%% file: paper.tex
\documentclass[10pt,conference]{IEEEtran}
\IEEEoverridecommandlockouts

\usepackage{amsmath,amssymb,amsfonts}
\usepackage{algorithmic}
\usepackage[pdftex]{graphicx}
\usepackage{grffile}
\usepackage{textcomp}
\usepackage{xcolor}
\usepackage[numbers]{natbib}
\usepackage{subcaption}
\usepackage{xspace}
\usepackage{booktabs, tabularx}
\usepackage{url}
\usepackage{multirow}
\usepackage[colorlinks=false, linkcolor=blue, citecolor=blue, bookmarks=true,pagebackref=false]{hyperref}
\usepackage{tcolorbox}
\usepackage{verbatim}
\usepackage{balance}
\usepackage{float}
\usepackage[T1]{fontenc}
\usepackage{listings}
\usepackage{makecell}

\input{setup}

\usepackage{hyperref}
\hypersetup{
  pdfauthor={Anonymous},
  pdftitle={},
  pdfsubject={},
  pdfkeywords={}
}

\begin{document}

\title{Screencast-Based Analysis of User-Perceived GUI Responsiveness}

\author{
    \IEEEauthorblockN{Wei Liu\textsuperscript{1}, Linqiang~Guo\textsuperscript{1}, Yi Wen Heng\textsuperscript{1}, Chenglin Li\textsuperscript{1},Tse-Hsun (Peter) Chen\textsuperscript{1}, Ahmed E. Hassan\textsuperscript{2}}
    \IEEEauthorblockA{\textit{\textsuperscript{1}Software PErformance, Analysis, and Reliability (SPEAR) lab, Concordia University, Montreal, Canada}}
    \IEEEauthorblockA{\textit{\textsuperscript{2}Queen's University, Canada}}

    \IEEEauthorblockA{w\_liu201@encs.concordia.ca, g\_linqia@live.concordia.ca, he\_yiwen@encs.concordia.ca}
    \IEEEauthorblockA{chenglin.li@mail.concordia.ca, peterc@encs.concordia.ca, Ahmed@cs.queensu.ca}
}

\maketitle

\pagenumbering{arabic}
\pagestyle{plain}
\begin{abstract}

GUI responsiveness is critical for a positive user experience in mobile applications. Even brief delays in visual feedback can frustrate users and lead to negative reviews. 
However, detecting and quantifying such user-perceived delays remains challenging, especially in industrial testing pipelines that evaluate thousands of apps daily across diverse devices and OS versions. Existing techniques based on static analysis or system metrics, while useful, may not accurately capture user-perceived issues or scale effectively.  

In this experience paper, we present \tool, a lightweight and black-box technique that measures GUI responsiveness directly from mobile screencasts---video recordings captured during automated GUI testing. 
\tool detects user interactions and visual delays, helping developers identify GUI performance issues that affect the user experience. 
It uses computer vision to detect user interactions and analyzes frame-level visual changes to compute two key metrics: response time (from user action to first visual feedback) and finish time (until visual feedback stabilizes). 
We evaluate \tool on a manually annotated benchmark of 2,458 interactions from 64 popular Android apps. \tool achieves 0.96 precision and 0.93 recall in detecting interactions, and measures response
and finish times within 50\,ms and 100\,ms error, respectively, for over 89\% of interactions. The tool has been deployed in an industrial testing pipeline and analyzes thousands of screencasts daily, uncovering responsiveness issues missed by traditional tools and improving performance debugging efficiency.

\end{abstract}

\begin{IEEEkeywords}
GUI responsiveness, mobile apps, mobile performance, user experiences
\end{IEEEkeywords}

\input{texFiles/intro}
\input{texFiles/background}
\input{texFiles/related}
\input{texFiles/methodology}

\input{texFiles/evaluation}

\input{texFiles/rq1}
\input{texFiles/rq2}

\input{texFiles/rq3}
\input{texFiles/rq4}

\input{texFiles/rq5}
\input{texFiles/discussion}

\input{texFiles/threats}
\input{texFiles/conclusion}

\balance

\footnotesize
\bibliographystyle{IEEEtranN}
\bibliography{IEEEabrv,ref}

\end{document}

%% file: setup.tex
\newcommand{\REM}[1]{}

\usepackage{xcolor}
\definecolor{dkgreen}{rgb}{0,0.6,0}
\definecolor{gray}{rgb}{0.5,0.5,0.5}
\definecolor{mauve}{rgb}{0.58,0,0.82}
\definecolor{dkred}{rgb}{0.6, 0.1, 0}
\definecolor{dkblue}{rgb}{0, 0, 0.6}
\definecolor{lightgreen}{RGB}{205, 255, 216}
\definecolor{lightred}{RGB}{255, 220, 224}
\definecolor{lightblue}{RGB}{179, 217, 255}
\definecolor{brass}{rgb}{0.71, 0.65, 0.26}
\definecolor{violet(ryb)}{rgb}{0.53, 0.0, 0.69}
\definecolor{mediumorchid}{rgb}{0.73, 0.33, 0.83}
\definecolor{darkorange}{rgb}{1.0, 0.55, 0.0}
\definecolor{navyblue}{rgb}{0.36, 0.54, 0.66}
\definecolor{amber}{rgb}{1.0, 0.75, 0.0}

\usepackage[linesnumbered,ruled,vlined]{algorithm2e}
\usepackage{amsmath}\newcommand{\tabincell}[2]{\begin{tabular}{@{}#1@{}}#2\end{tabular}}

\newcommand{\phead}[1]{\vspace{1mm} \noindent {\bf #1}}

\newcommand{\tool}{{{MobileGUIPerf}}\xspace}
\newcommand{\rqbox}[1]{\begin{tcolorbox}[left=4pt,right=4pt,top=4pt,bottom=4pt,colback=gray!5,colframe=gray!40!black,before skip=2pt,after skip=2pt]#1\end{tcolorbox}}

%% file: texFiles/intro.tex
\section{Introduction}
\label{sec:introduction}

Mobile device manufacturers must ensure high-quality user experiences across a diverse and large set of apps. To maintain user satisfaction, they must continuously perform  large-scale testing by executing automated Graphical User Interface (GUI) tests on thousands of apps daily across multiple device models and software versions. 
In this setting, detecting performance issues, especially those related to GUI responsiveness from the user's perspective, is both critical and technically challenging.

GUI responsiveness plays a crucial role in the user experience of mobile applications. Even brief delays in visual feedback after tapping, such as sluggish button responses, can frustrate users and make the app feel unusable. Poor responsiveness is among the top reasons users abandon apps or leave negative reviews~\cite{khalid2014mobile}, making it a critical factor in perceived app quality. 
However, despite its importance, accurately detecting and measuring these issues at scale remains a challenge in current industrial testing workflows.


Existing studies on mobile performance typically rely on static or dynamic analysis of source code~\cite{liu2014characterizing, afjehei2019iperfdetector, 2019_SANER_Characterizing_and_Detecting_Inefficient_Image_Displaying_Issues, 2020_EMSE_statically_detectable_performance_issues, 2021_ICSE_IMGDroid_Detecting_Image_Loading_Defects, 2023_ASE_Detection_Thread_Misuses} or system-level metrics~\cite{2012_OSDI_AppInsight, 2019_MOBILESoft_PerfProbe, 2022_EMSE_AppSPIN, adb, perfetto}. While these techniques provide valuable insights, they suffer from two major limitations. First, they often require access to source code or modifications to application binaries, making them difficult to apply in large-scale settings. This limitation is further exacerbated by the frequent updates of mobile apps, which demand continual instrumentation and maintenance. 
Second, and more importantly, these techniques do not capture performance from the user’s perspective. As a result, many of the issues they detect may not be noticeable by users and have minimal impact on perceived experience~\cite{2022_IST_resource_influences_UI_responsiveness}. 

To address these limitations, we collaborated with our industry partner and developed \tool, a lightweight and black-box technique for measuring GUI responsiveness directly from the user's perspective. Since recording video screencasts of app usage is a common practice in automated GUI testing~\cite{2024_EMSE_characteristics_of_visual_issue_reports}, \tool leverages these recordings to assess responsiveness. 
It employs computer vision techniques to automatically 1) identify user interactions and 2) measure their responsiveness in terms of response and finish times. 
By analyzing screencasts, which faithfully reflect what users actually see, \tool enables developers to detect user-perceived performance issues. 

\tool first applies a computer vision-based object detection model to identify visual tap indicators in screencasts, segmenting the video screencast into user interactions. These tap indicators are generated by Android's built-in \texttt{Show taps} feature, which highlights user actions such as taps by displaying a visual circle at the point of contact. \tool then uses frame differencing techniques based on the Structural Similarity Index Measure (SSIM)~\cite{2004_SSIM_Transactions_on_Image_Processing} to detect visual changes and compute GUI responsiveness. For each interaction, it computes two key metrics: the response time (the time from tap to the first visible change) and the finish time (the time until the visual feedback stabilizes). 

To evaluate \tool, we constructed a benchmark dataset consisting of 2,458 user interactions from 64 of the most popular mobile apps. Each interaction was manually analyzed and annotated with the corresponding response and finish times. 
Our results show that \tool achieves a precision of 0.96 and a recall of 0.93 in identifying user interactions from screencasts. For these interactions, \tool measures GUI responsiveness with high accuracy: 95\% of interactions have measurement errors within 3 frames (50\,ms) for response time, and 89\% fall within 6 frames (100\,ms) for finish time.
Additionally, \tool is highly efficient, processing a 5-second screencast in approximately 9 seconds.  
The system-level recording overhead introduced by video capture is minimal--16\,ms for response time and 51\,ms for finish time.
The tool has been integrated into our industry partner's automated testing pipeline and now analyzes thousands of recordings daily.

The main contributions of this experience paper are as follows: 
\begin{itemize}
    \item We release a manually annotated dataset containing 2,458 user interactions, each labeled with the corresponding response and finish times, to encourage future research on this important topic. The dataset is publicly available at \url{https://anonymous.4open.science/r/gui-response-2293/}.

	\item 
    We present \tool, the first black-box technique that automatically measures GUI responsiveness from screencasts, reflecting user-perceived performance. 


    \item \tool achieves high precision and recall in identifying user interactions triggered by user operations (e.g., taps or swipes), and accurately measures their GUI responsiveness in terms of response and finish times.

    \item \tool has been deployed in our industry partner's testing pipeline, where it analyzes thousands of screencasts daily and has helped uncover performance issues that traditional tools failed to detect. 


\end{itemize}


\phead{Paper organization.} 
Section~\ref{sec:background} discusses background on GUI responsiveness and the limitations of existing approaches. Section~\ref{sec:related} reviews related work. Section~\ref{sec:approach} details our approach. Section~\ref{sec:evaluation} evaluates our approach on real-world datasets, while Section~\ref{sec:discussion} discusses its industrial deployment and practical impact. 
Section~\ref{sec:threats} outlines threats to validity, and Section~\ref{sec:conclusion} concludes the paper.

%% file: texFiles/background.tex
\section{Background}
\label{sec:background}

\subsection{GUI Responsiveness}

Responsiveness in the Graphical User Interface (GUI) is a key non-functional property that greatly impacts user satisfaction. When a mobile application responds slowly to user actions, such as tapping a button or swiping a screen, it often appears unresponsive or laggy, which can frustrate users and even lead to app abandonment or negative reviews~\cite{khalid2014mobile, liu2014characterizing, afjehei2019iperfdetector}. Table~\ref{fig_responsiveness_metrics} summarizes common GUI responsiveness metrics based on findings in human-computer interaction (HCI) and performance research~\cite{yan2012fast, 10.1145/1062745.1062747}. \textit{Response time} refers to the time between a user action and the first visible frame update in the GUI. Prior research in HCI suggests users can feel delays in responses that take more than 100\,ms~\cite{10.1145/1062745.1062747}. Hence, even minor delays in GUI updates may thus be perceived as unresponsiveness. 
\textit{Finish time} captures the duration from the user’s action to when the GUI completes all visual transitions (i.e., usually when the last frame stops changing) and becomes ready for the next user interaction. 

\begin{table}
\caption{GUI responsiveness metrics.}
\label{fig_responsiveness_metrics}
\centering
\scalebox{0.95}{
    \setlength{\tabcolsep}{3.5pt}
    \begin{tabular}{lp{6cm}}
    \toprule
    GUI responsiveness & Definition\\
    \toprule
    Response time & Duration from user input to the \textbf{first} visible GUI frame update. \\
    \midrule
    Finish time & Duration from user input to the \textbf{final} GUI frame update. \\
    
    \bottomrule
    \end{tabular}}
\end{table}

\subsection{Limitations of Existing Approaches}
However, measuring GUI responsiveness remains challenging in practice. Most existing tools rely on system-level metrics such as CPU usage, memory consumption, or UI thread activity~\cite{adb, perfetto, CPU_Profiler, 2022_IST_resource_influences_UI_responsiveness}. These techniques, while valuable from a system standpoint, have two major limitations:
\begin{enumerate}
    \item Limited Applicability: They often require source code access or intrusive instrumentation, which may not be feasible in black-box (e.g., testing third-party, closed-source apps for benchmarking) or large-scale testing scenarios (e.g., device manufacturers need to test hundreds of apps across many devices). 
    \item Lack of User Perspective: These tools monitor system-level metrics that do not always reflect what users perceive. For example, a UI thread may be busy working in the background, but does not produce any immediate visual changes on the screen~\cite{2022_IST_resource_influences_UI_responsiveness}. 
\end{enumerate}

\noindent As a result, developers are often left to manually inspect screencasts frame by frame—a labor-intensive process to understand how their app performs in real-world usage. 

The issue becomes more challenging in large-scale testing environments. For instance, our industry partner needs to run hundreds or even thousands of automated GUI tests across many third-party apps on various mobile devices daily, making it impossible to manually diagnose performance issues through screencasts. 
While existing tools like NightHawk~\cite{liu2022nighthawk} and OwlEyes~\cite{liu2020owl} (e.g., text overlap, misalignment) can detect visual UI issues, they primarily target display problems (e.g., text overlap and misalignment) rather than temporal aspects like UI unresponsiveness or delayed feedback. This gap highlights the need for an automated approach to assess GUI responsiveness directly from the user's visual perspective.

\subsection{Analyzing Screencasts as a User-Centric Solution}

\begin{figure*}[t]
	\centering
	\includegraphics[width=0.9\linewidth]{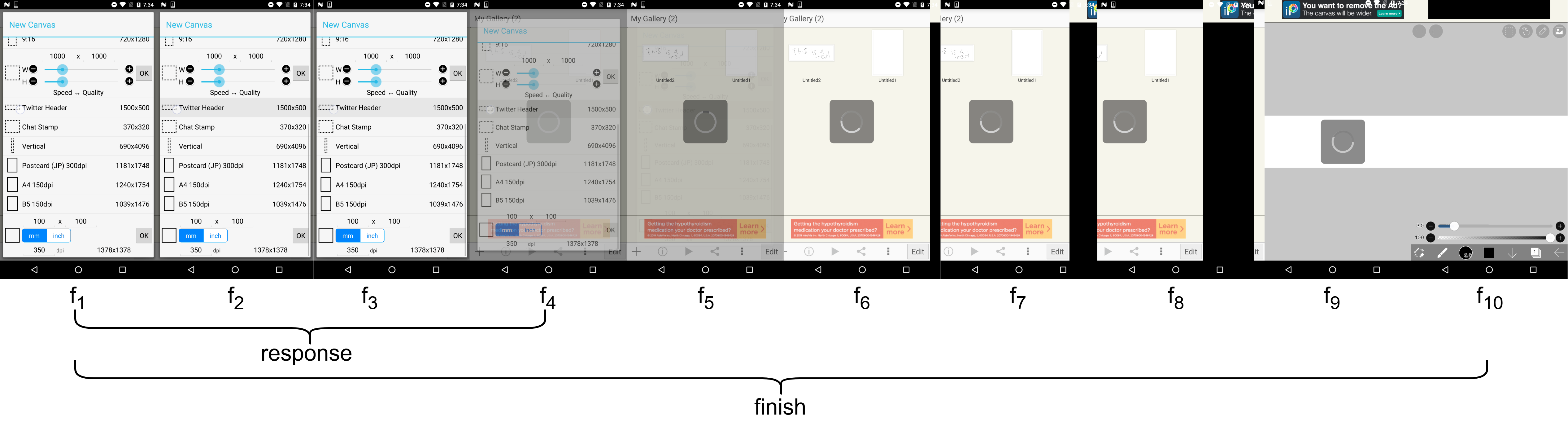}
	\caption{A sequence of frames in a recorded screencast, with each frame annotated by index (e.g., $f_{1}$). The response duration is from $f_1$ to $f_4$ while the finish duration is from $f_1$ to $f_{10}$.}
	\label{fig_reponsiveness}
\end{figure*}

To bridge this gap, we analyze screencasts, which are video recordings of mobile device screens that visually capture how the UI responds to user interactions. Screencasts are widely used in practice, particularly in industrial testing pipelines (e.g., for recording videos during automated testing) and bug reporting tools~\cite{2024_EMSE_characteristics_of_visual_issue_reports}. 
A screencast consists of a sequence of frames, each marked with a timestamp, providing a precise timeline of what users see.
Screencasts are often recorded at the same frame rate as the device~\cite{screen_record_frame_rate}. Each time the screen updates, a new frame is captured. For instance, most mobile devices typically operate at a standard frame rate of 60 Frames Per Second (FPS)~\cite{Android_frame_rate}, which is approximately 16.7 milliseconds~\cite{Android_Performance_Patterns} between consecutive frames. 
Each time the screen content is updated, a new frame is recorded, making screencasts a reliable source for analyzing the timing of UI responses.

\begin{figure*}
	\centering
	\includegraphics[width=0.85\linewidth]{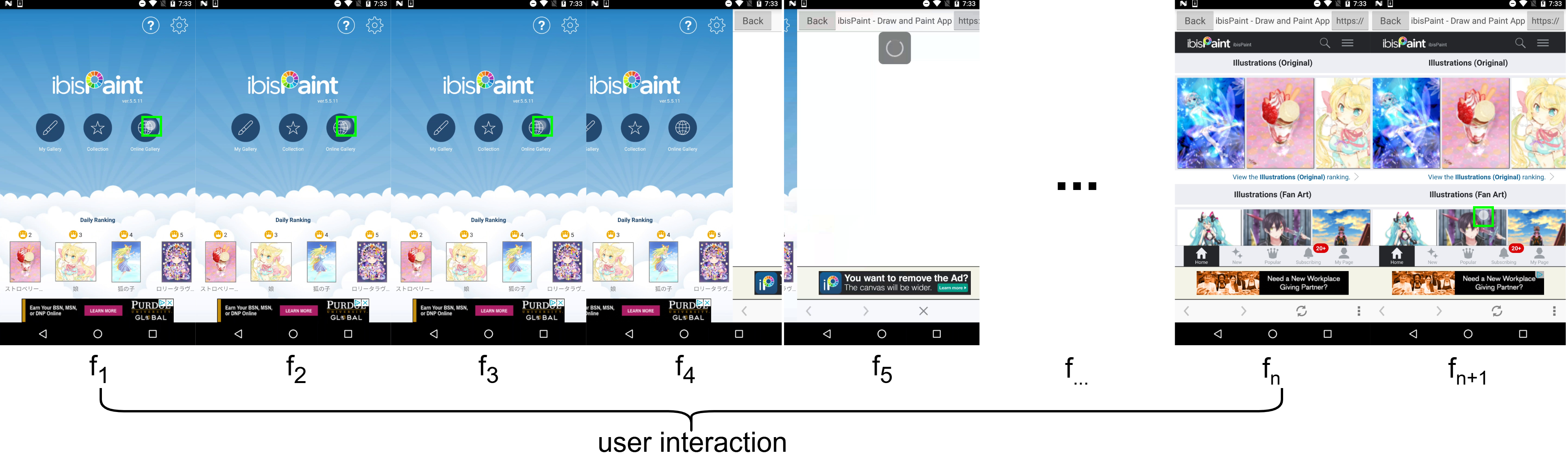}
	\caption{Illustration of user interaction segmentation. The tap indicator (shown at $f_1$, $f_2$, $f_3$, and $f_{n+1}$) is marked as green rectangle. The first user interaction starts at frame $f_1$ and ends at frame $f_{n}$. The next interaction starts at frame $f_{n+1}$.}
	\label{fig_screencast}
\end{figure*}

Figure~\ref{fig_reponsiveness} illustrates an example screencast composed of 10 frames, each labeled with a timestamp indicating its Presentation Time Stamp (PTS). For instance, frame 1 ($f_1$) is shown at timestamp 0 ms, followed by frame 2 ($f_2$) at timestamp 16 ms, and so on. In this scenario, the user taps a UI element called ``Twitter Header'' in the setting screen at $f_1$ to create a canvas for drawing. The mobile system begins responding at frame $f_4$. The system then transitions to the canvas and loads the drawing tools, completing the interaction by $f_{10}$. As a result, the response time is calculated as $f_4$ - $f_1$, and the finish time is $f_{10}$ - $f_1$.

In our collaboration with Company A, screencasts are commonly collected during automated test runs and crowd-sourced bug submissions, often with the Android ``Show taps'' feature enabled. This option overlays a semi-transparent circle at the user contact point. As shown in Figure~\ref{fig_screencast}, the circle provides a visual identification of user inputs and allows developers to analyze GUI responsiveness without internal app instrumentation.

This frame-level analysis enables accurate measurement of GUI responsiveness. Since every UI update is recorded, developers can determine exactly when the system begins to react and when it stabilizes, without relying on source code access or system-level instrumentation. By analyzing screencasts, developers can automate the evaluation of GUI responsiveness and detect issues that directly impact user experience. This makes screencasts especially useful for black-box testing and performance regression analysis in continuous integration (CI) workflows.

Despite their potential, there are challenges in automatically analyzing GUI responsiveness from screencasts. Accurately detecting the start and end of visual feedback is difficult, especially with subtle transitions or partial UI updates. Noise from background animations and varied app behaviors further complicates analysis~\cite{2023_ICSE_Efficiency_Matters_Speeding_Up_Automated_Testing, liu2025guiwatcherautomaticallydetectinggui}. Hence, we collaborate with our industry partner to address these challenges, enabling automated screencast analysis at scale. 

%% file: texFiles/related.tex
\section{Related Work}
\label{sec:related}
In this section, we review prior work related to our study. 

\phead{Mobile Performance Analysis.} 
Tools such as Android Lint~\cite{Android_Lint}, FindBugs~\cite{FindBugs}, PMD~\cite{PMD}, and Infer~\cite{Infer} have been widely used to detect performance issues in mobile apps. These tools typically use static analysis to identify performance anti-patterns or inefficient code structures. Prior work has extended these techniques to identify source-level performance problems~\cite{liu2014characterizing, APEChecker_ASE_2018, afjehei2019iperfdetector, 2023_ASE_Detection_Thread_Misuses}. However, such tools often miss user-perceived delays and depend heavily on predefined rules, which limits their applicability in real-world scenarios. 

Profiling tools such as Android Debug Bridge (adb)~\cite{adb}, Perfetto~\cite{perfetto}, and Android Studio Profiler~\cite{Android_Studio_Profiler} monitor metrics like CPU usage, memory consumption, and GPU activity. However, the performance metrics these tools monitor do not always reflect the responsiveness issues perceived by end users~\cite{2022_IST_resource_influences_UI_responsiveness}. Dynamic instrumentation techniques (e.g., AppInsight~\cite{2012_OSDI_AppInsight}, PerfProbe~\cite{2019_MOBILESoft_PerfProbe}, AppSPIN~\cite{2022_EMSE_AppSPIN}) collect detailed runtime events, but they still operate at the system level and may overlook visual delays. In contrast, our approach directly analyzes UI behavior through screencasts, enabling the detection of responsiveness issues from the end user's perspective.

\phead{Analysis of mobile app screencasts.} 
Previous work has analyzed mobile screencasts to support various testing and debugging tasks, such as translating video recordings into replayable scenarios~\cite{2020_ICSE_translating_video_recordings_of_mobile_app_usages, 2023_TSE_Translating_Video_Recordings_of_Complex_Mobile_App_UI_Gestures}, detecting janky frames~\cite{liu2025guiwatcherautomaticallydetectinggui}, and identifying advertisements during app testing~\cite{guo2024popsweeperautomaticallydetectingresolving}, and automatically replaying visual bug reports for Android apps~\cite{GIFdroid}.
One related technique, AdaT~\cite{2023_ICSE_Efficiency_Matters_Speeding_Up_Automated_Testing}, aims to accelerate automated testing by detecting when a UI transition stabilizes. Like our method, AdaT uses computer vision techniques to classify the rendering state of frames. In contrast, \tool focuses on detecting user interactions and localizing keyframes to measure GUI responsiveness. 

While prior techniques offer valuable insights, they often operate at the code or system level or serve different goals, leaving a gap in capturing user-perceived responsiveness---precisely the focus of our work.

%% file: texFiles/methodology.tex
\section{Approach}
\label{sec:approach} 

\begin{figure*}[t]
	\centering
	\includegraphics[width=0.85\linewidth]{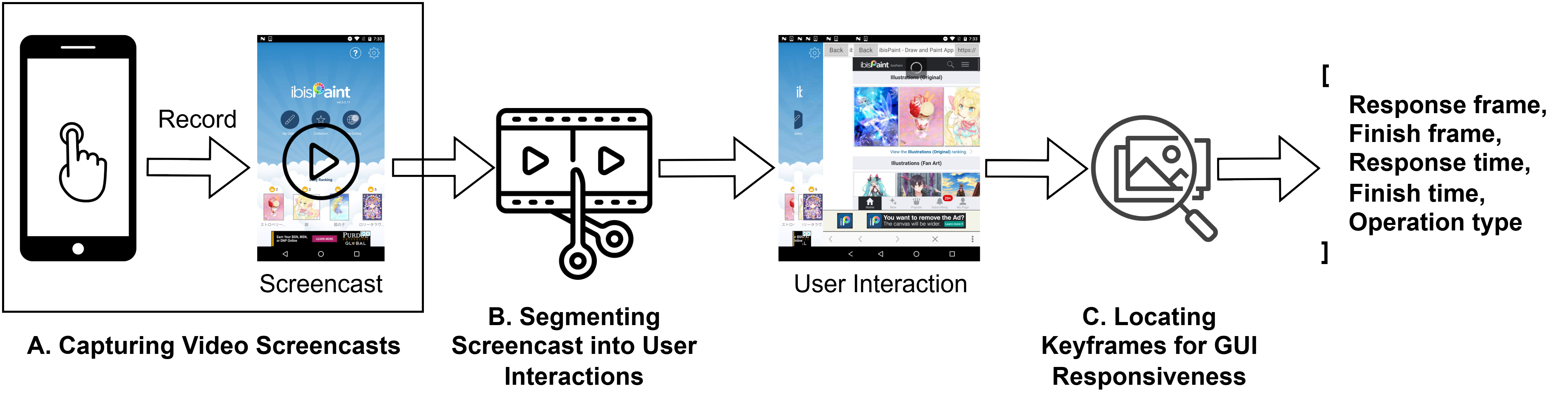}
	\caption{The overall architecture of \tool.}
	\label{fig_overview_approach}
\end{figure*}

We propose \tool, an automated framework developed in collaboration with Company A, to measure GUI responsiveness from the end-user perspective by analyzing mobile screencasts. \tool supports large-scale app testing pipelines by computing the response and finish time for each user interaction without requiring source code or instrumentation. As illustrated in Figure~\ref{fig_overview_approach}, \tool consists of three main components: (1) capturing video screencasts, (2) segmentation of screencast into user interactions, and (3) locating keyframes for GUI responsiveness. Algorithm~\ref{alg:mobileguiperf} shows the specific steps and design details of \tool. The framework leverages computer vision techniques to identify user interactions and locate keyframes that indicate the response and finish of user interactions. Importantly, \tool operates without access to source code or instrumentation, making it suitable for black-box testing and scalable performance analysis.

\begin{algorithm}[t]
\caption{\tool: Measuring GUI Responsiveness from Screencasts.}
\label{alg:mobileguiperf}
\KwIn{Screencast video $V$ recorded at $f$ FPS}
\KwOut{List of user interactions $\mathcal{I}$ with response and finish times}

\textbf{Step 1: Frame Extraction} \\
Extract all frames $\mathcal{F} = \{f_1, f_2, ..., f_n\}$ from $V$ with timestamps.\\

\textbf{Step 2: Tap Indicator Detection} \\
Use Faster R-CNN to detect tap indicators in each frame.\\
Group frames into segments $\mathcal{S} = \{s_1, s_2, ..., s_m\}$ based on tap indicator intervals.

\textbf{Step 3: User Interaction Classification} \\
\ForEach{segment $s_i \in \mathcal{S}$}{
    Track centroid of tap indicator across frames in $s_i$\; 
    \uIf{indicator is static}{
        $type_i \gets$ Tap\;
    }
    \uElse{
        $type_i \gets$ Swipe\;
    }
}

\textbf{Step 4: Keyframe Detection (Per Interaction)} \\
\ForEach{segment $s_i \in \mathcal{S}$}{
    Compute frame-to-frame visual similarity scores\; 
    Apply Isolation Forest to detect outlier frames\; 
    Identify response frame $f_{resp}$ and finish frame $f_{fin}$ based on visual change points\;

    Compute:
    \begin{itemize}
        \item Response time: $RT_i = t(f_{resp}) - t(f_{start})$
        \item Finish time: $FT_i = t(f_{fin}) - t(f_{start})$
    \end{itemize}
    
    Store interaction $\mathcal{I}_i = \langle f_{start}, RT_i, FT_i, type_i \rangle$
}

\Return $\mathcal{I} = \{\mathcal{I}_1, \mathcal{I}_2, ..., \mathcal{I}_m\}$

\end{algorithm}

\subsection{Capturing Video Screencasts}
Since video recording can introduce performance overhead, we minimize this impact by using the open-source tool scrcpy~\cite{scrcpy}, which captures the device’s screen via efficient video streaming. Compared to the system’s built-in recorder, scrcpy introduces significantly lower CPU usage and latency, enabling a more realistic reflection of the app’s actual performance. Scrcpy works by streaming the device’s screen content over USB to a host machine, where the video is encoded and recorded. This process offloads the processing from the mobile device to external machines, thereby reducing runtime overhead. 
We record screencasts at the device’s full frame rate (typically 60 FPS) to ensure that all visual changes are preserved with high temporal precision. A higher frame rate reduces the interval between consecutive frames (around 16.6 ms), capturing finer-grained visual transitions and enabling more accurate measurement of responsiveness. Once recorded, the screencasts are saved as video files for subsequent offline analysis.

\subsection{Segmenting Screencast into User Interactions}

To analyze GUI responsiveness at the level of individual actions, \tool segments each screencast into separate \textbf{user interactions}, i.e., sequences of frames that correspond to a single user action. This stage consists of two steps: identifying user interaction boundaries and inferring the type of user action. 

\subsubsection{Identifying User Interaction Boundaries}

We segment screencasts into user interactions by leveraging Android's \texttt{Show taps} option when recording. As shown in Figure~\ref{fig_screencast}, this feature overlays a semi-transparent circle at the point of contact when a user touches the screen. The circle appears under the finger, follows its movement, and typically remains visible for several consecutive frames before gradually fading. Its duration and opacity vary depending on the type of interaction (e.g., tap, swipe). While originally intended for debugging and screen recording~\cite{record_android, configure_android}, this feature has been adopted in prior work on UI analysis and recording-and-replay~\cite{2020_ICSE_translating_video_recordings_of_mobile_app_usages, 2023_TSE_Translating_Video_Recordings_of_Complex_Mobile_App_UI_Gestures}. In our study, these tap indicators serve as visual cues for segmenting the screencast into individual user interactions.

We define a \textbf{user interaction} as a sequence of frames that begins with a frame showing the onset of a tap indicator and ends just before the start of the next indicator (i.e., the next user action). As illustrated in Figure~\ref{fig_screencast}, a tap indicator appears at frame 1, remains visible in frames 2 and 3, and disappears in frame 4. If the next tap appears at frame n+1, the frames from 1 to n are considered a single user interaction. Since the tap indicator only appears for a very short time on a few frames before the end of GUI reactions, users typically tap the screen after the previous indicator has already disappeared.

To identify interaction boundaries, we first extract all frames from the screencast and apply a pre-trained object detection model~\cite{2020_ICSE_translating_video_recordings_of_mobile_app_usages} based on Faster R-CNN~\cite{Faster-R-CNN} to detect the presence and position of tap indicators. This model was trained by prior work~\cite{2020_ICSE_translating_video_recordings_of_mobile_app_usages} on a synthetic dataset comprising 15,000 UI images, generated by superimposing touch indicators with varying opacity levels onto 5,000 unique Android app screenshots. This large-scale training corpus enables the model to generalize well across diverse mobile interfaces. Because tap indicators are often visually fused with app content, traditional image processing methods can produce suboptimal results. In contrast, the Faster R-CNN--based detector achieves robust performance across varied UI contexts. Since a single user operation produces one tap indicator that typically persists across multiple consecutive frames, we group consecutive detections into tap sequences and segment the screencast into user interactions based on the first frame in each sequence.

\subsubsection{Classifying the Type of User Action}
To contextualize the measured responsiveness, we classify each user interaction as either a \textbf{Tap} or a \textbf{Swipe}. This binary classification aligns with our focus on GUI responsiveness metrics: Tap interactions typically involve a brief, stationary touch, where both response time (i.e., how quickly the GUI reacts) and finish time (i.e., how long until the GUI finishes reacting) are meaningful. In contrast, Swipe interactions encompass continuous gestures---such as scrolling, swiping, or drawing---where response time is more critical, and a longer finish time may reflect expected behavior rather than performance issues.

Building on the previously segmented user interactions, we analyze the spatial movement of the tap indicator within each interaction to determine its type. The indicator appears as a semi-transparent circle that follows the user’s finger throughout the interaction. As a result, its motion provides a visual proxy for the user’s gesture. Specifically, we compute the center point $(x, y)$ of the bounding box enclosing the tap indicator in the first and last frames where it appears. If the movement---measured as the Euclidean distance between these two points---is less than 10 pixels, we classify the interaction as a Tap; otherwise, it is classified as a Swipe. This threshold accounts for minor detection noise in the tap indicator's position, even during stationary touches. We empirically selected 10 pixels based on visual inspection and validation against a subset of annotated interactions.

\subsection{Locating Keyframes for GUI Responsiveness} 
After segmenting the screencast into user interactions, the next step is to locate two keyframes for each interaction segment: the \textit{response frame} ($f_{\text{resp}}$), where the first visual feedback appears, and the \textit{finish frame} ($f_{\text{fin}}$), where the GUI becomes visually stable. We can calculate the following two metrics to measure the GUI responsiveness: 
$\text{Response Time as }  t(f_{\text{resp}}) - t(f_{\text{start}}) \text{ and }
\text{Finish Time as }  t(f_{\text{fin}}) - t(f_{\text{start}})$, 
where $t(f)$ denotes the timestamp of frame $f$, and $f_{\text{start}}$ is the first frame of the interaction.

To identify these keyframes, we first compute visual similarity scores between consecutive frames within each user interaction. Given a segment $\mathcal{S} = \{f_1, f_2, \ldots, f_n\}$, we calculate:

\[
\Delta = \left\{ \text{sim}(f_j, f_{j+1}) \mid j = 1, 2, \ldots, n-1 \right\},
\]
\noindent where $\text{sim}(f_j, f_{j+1})$ is the structural similarity index (SSIM) between frames $f_j$ and $f_{j+1}$. SSIM captures structural changes in the image and is more robust to color or resolution differences than raw pixel comparisons~\cite{2004_SSIM_Transactions_on_Image_Processing}.

We then apply the Isolation Forest algorithm~\cite{liu2008isolation} to the sequence $\Delta$ to identify anomalous frames that reflect significant visual transitions. Isolation Forest is an unsupervised anomaly detection technique that isolates outliers using random binary partitions, which has been shown to be efficient and accurate~\cite{liu2008isolation}. 
In our context, frames with large visual differences are flagged as anomalies. We select the first anomaly as the \textit{response frame} and the last as the \textit{finish frame}, corresponding to the start and finish of visible GUI feedback. To improve accuracy, we incorporate domain knowledge into the anomaly detection process. Specifically, industry experts noted that end users typically associate responsiveness with substantial visual transitions (e.g., navigating to a new screen), rather than minor effects such as a button dimming. Based on this insight, we introduce an offset in the Isolation Forest algorithm to refine the position of the response frame. This adjustment helps align the detection with user expectations of responsiveness. Finally, we extract timestamps embedded in the video file to compute response and finish times based on the identified response and finish frames.

\subsection{A Working Example of \tool}

\begin{figure}
	\centering
	\includegraphics[width=0.95\linewidth]{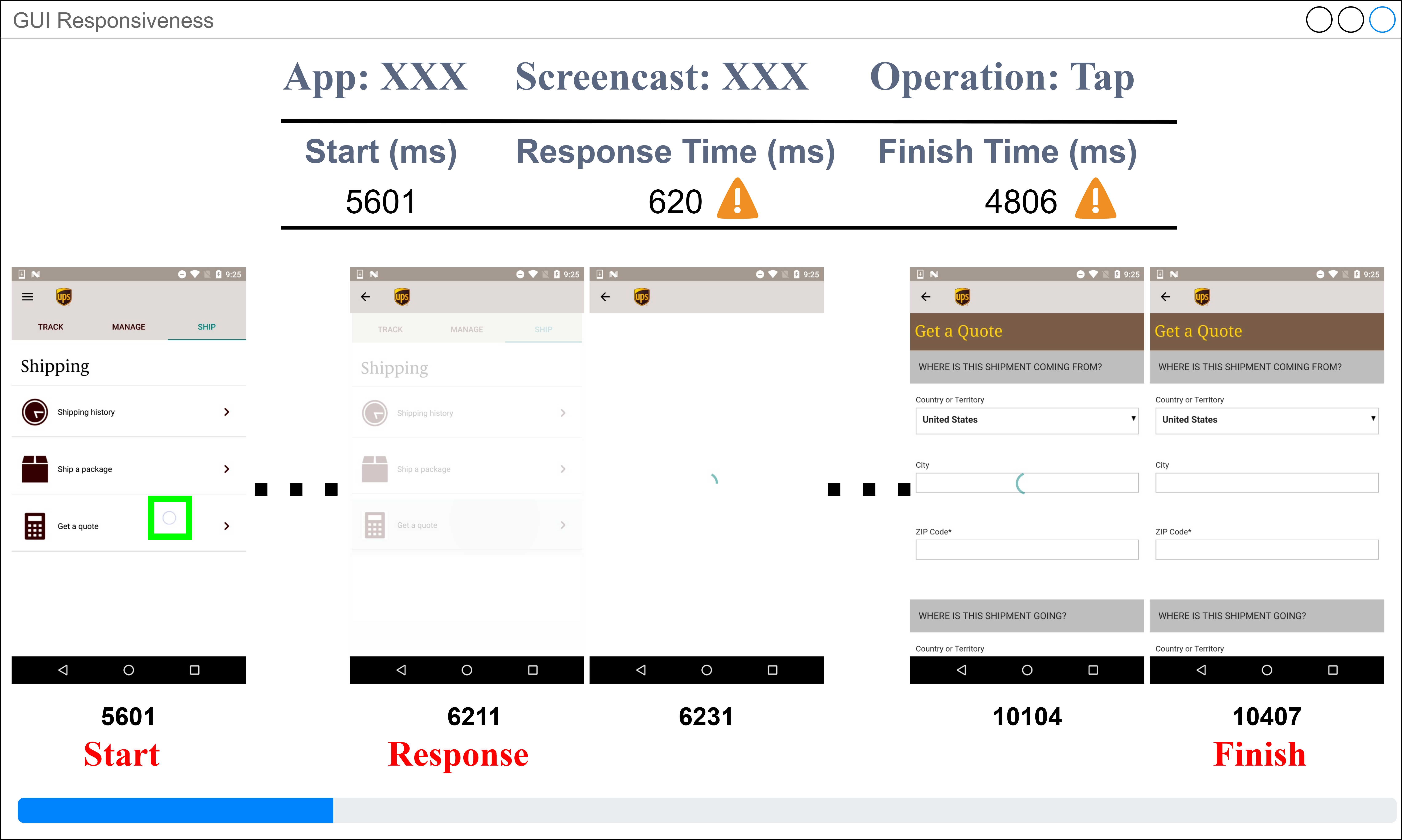}
	\caption{An illustrative interface of \tool.}
	\label{fig_interface}
\end{figure}

To illustrate how \tool operates in practice, we present an end-to-end example of its output.
\tool is deployed in production and processes thousands of screencasts daily. It enables accurate and scalable measurement of GUI responsiveness directly from screencasts, without requiring access to source code, instrumentation, or prior knowledge of the app's internal logic---making it well suited for black-box and large-scale testing. Figure~\ref{fig_interface} presents an illustrative example of the output generated by \tool. Due to the non-disclosure agreement, we cannot show the actual production interface, but the figure reflects a representative report format. Each report corresponds to a detected user interaction in the screencast and includes response and finish times, along with severity indicators (whether it exceeds certain thresholds). Developers can directly analyze the videos frame by frame and can use this information to understand how the interaction was triggered and assess its impact based on severity, enabling efficient prioritization and debugging. 

%% file: texFiles/evaluation.tex
\section{evaluation} 
\label{sec:evaluation}
\subsection{Experimental Setup}
\label{sec:setup}
To evaluate the accuracy of \tool in measuring GUI responsiveness, we require a dataset of screencast videos recorded during the real-world usage. While some prior studies provide mobile app recordings, existing datasets have significant limitations. For instance, the RICO~\cite{RICO} dataset consists of user interaction recordings in GIF format, sampled at one frame per second (i.e., 1 FPS), which is insufficient for fine-grained video analysis. 
Hence, we manually annotate the GUI responsiveness (i.e., response and finish time) based on the videos in the V2S dataset~\cite{2020_ICSE_translating_video_recordings_of_mobile_app_usages}. The dataset contains 128 videos collected from 64 popular Android applications on Google Play (top two from each of 32 app categories), such as Amazon, LinkedIn, and Airbnb. Each video was recorded with visible tap indicators overlaid on the mobile screen to illustrate user actions such as taps and swipes at a high FPS. 

One limitation is that the V2S dataset does not provide frame-level annotations indicating the start, response, or finish of each user interaction. To construct a high-quality ground truth, three authors of this paper, each with over 5 years of mobile testing experience, independently annotated the video. Each annotator independently examines every frame to identify the start frame, response frame, and finish frame of every user interaction. For each user interaction, we recorded a 5-tuple: $[f_\text{start}, f_\text{response}, f_\text{finish}, f_\text{end}, \textit{type}]$, where $f_\text{end}$ is computed as the last frame before the next interaction, and $\textit{type}$ denotes the user action type (e.g., tap or swipe). After the initial annotation, the annotators met to resolve any discrepancies and reach a consensus. These annotations were later used to evaluate the accuracy of our automated interaction detection approach. 

\begin{figure}
	\centering
    \includegraphics[width=\linewidth]{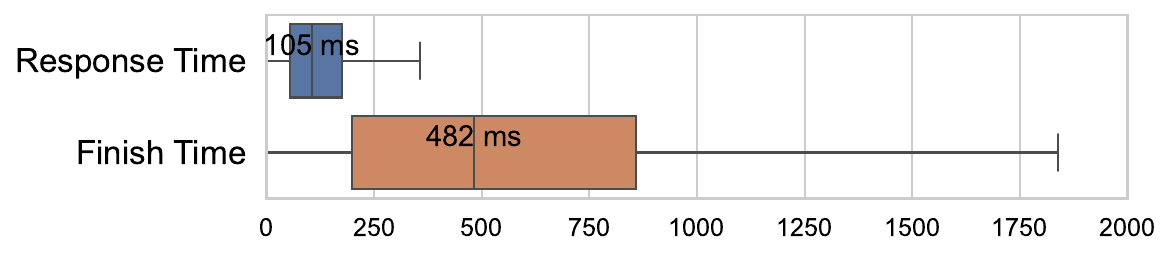}
	\caption{Distribution of GUI responsiveness times (ms). The median response time is 105\,ms, and the median finish time is 482\,ms.}
	\label{fig_dataset_distribution}
\end{figure}

In total, we collected 2,458 GUI-based user interactions from these 128 videos, with an average of 38 actions per mobile app. Figure~\ref{fig_dataset_distribution} illustrates the distribution of response and finish times across these interactions. Overall, the median response and finish times are 105\,ms and 482\,ms, respectively.

%% file: texFiles/rq1.tex
\subsection*{RQ1: What is the Accuracy of User Interaction Detection?}
\label{sec:rq1}

A mobile screencast often records multiple user interactions performed during a GUI test. To measure responsiveness at the user interaction level, our framework first segments the screencast into individual user actions. In this RQ, we evaluate how accurately \tool can identify and segment these interactions from the screencast.

\phead{Approach.} We use the manually annotated dataset (described in Section~\ref{sec:setup}) as the ground truth. In the ground truth, each user interaction is annotated as $[f_\text{start}, f_\text{response}, f_\text{finish}, f_\text{end}, \textit{type}]$. In our evaluation, a user interaction detected by \tool is considered \textit{correctly identified} if and only if:
\begin{enumerate}
   \item It matches the same user operation type as the corresponding ground truth interaction (e.g., tap, swipe), and
   \item It has the same start frame ($f_{start}$) as the ground truth annotation.
\end{enumerate}
Since each user operation in the screencast corresponds to a single user interaction, both the predicted and ground truth interactions must represent only one operation.

\phead{Metrics.} To evaluate the accuracy of \tool in identifying user interactions, we use the classic evaluation metrics:  precision, recall, and F1-score. Precision is the proportion of correctly identified user interactions among all identified user interactions, recall is the proportion of correctly identified user interactions among all actual user interactions, and F1-score is the harmonic mean of precision and recall. 

\phead{Results.} Table~\ref{tab:resutls_identify_interactions} shows the results of \tool in identifying user interactions across mobile apps. The results are computed based on all detected and ground-truth interactions across the dataset. The tool achieves a precision of 0.96 and a recall of 0.93, demonstrating high accuracy in extracting interactions from screen recordings.

Although we achieved high precision and recall, we still observe that some user interactions are incorrectly identified. We manually examined the results and found that most of them were caused by the incorrect detection of taps. Certain UI elements may be visually similar to tap indicators, resulting in incorrect user interactions. In addition, these misclassifications may interfere with segmentation boundaries, causing valid interactions to be incorrectly split or missed altogether.

\begin{table}[t]
	\centering
	\caption{Accuracy of \tool in identifying user interactions. }
	\label{tab:resutls_identify_interactions}
	\scalebox{0.95}{
         \setlength{\tabcolsep}{9mm}

	\begin{tabular}{rrr}
		\toprule
		\tabincell{r}{Precision} & \tabincell{r}{Recall} & \tabincell{r}{F1-Score} \\
		\midrule
		0.96 & 0.93 & 0.94 \\
		\bottomrule
	\end{tabular}}
\end{table}

\rqbox{\tool can detect user interactions accurately from screencasts, achieving a precision of 0.96 and a recall of 0.93. }

%% file: texFiles/rq2.tex
\subsection*{RQ2: How Accurate is \tool in Measuring GUI Responsiveness?}
In this RQ, we assess how accurately \tool measures the responsiveness (i.e., response time and finish time) of successfully identified interactions. 

\phead{Approach.} We use two metrics to quantify measurement accuracy: 
\begin{itemize}
	\item \textbf{Response time accuracy.} We compute the \textit{mean absolute error} (MAE) averaged across all user interactions. For each interaction, we calculate the absolute difference between the extracted and actual response time, measured in milliseconds, and the number of frames.

    \item \textbf{Finish time accuracy.} Similarly, the accuracy of finish time is measured using the \textit{mean absolute error} (MAE), defined as the absolute difference between the extracted and actual finish times, also measured in milliseconds, and the number of frames.
\end{itemize}

As mentioned above, we report the MAE for two aspects: 1) time in milliseconds and 2) video frames. We report frame-based and millisecond-based accuracy to provide a more comprehensive evaluation. Since \tool operates at the frame level, frame-based accuracy provides a direct view of precision, independent of the recording frame rate. In contrast, millisecond-based errors can vary depending on the frame rate of each video. Hence, presenting two metrics provides a detailed evaluation of \tool under different conditions. 

\phead{Results.} Table~\ref{tab:response_time_errors} presents the distribution of response time measurement accuracy by \tool. At the frame level, 92\% of the user interactions match exactly with the ground truth (0-frame difference). Furthermore, 95\% of the user interactions are within a 3-frame margin. In terms of the actual difference in time, the MAE is only 33 ms, with over 94\% of the cases having a time difference of less than 50 ms. Based on practitioner feedback, the results are widely acceptable due to their high accuracy and minimal time differences, as response times less than 100 ms are often imperceptible to users~\cite{1968_AFIPS_Response_time_in_man_computer, 1994_Usability_Engineering, 2015_High_Performance_Android, 2022_TSE_Survey_of_Performance_Optimization}. Table~\ref{tab:finish_time_errors} presents the distribution of finish time measurement accuracy produced by \tool. At the frame level, \tool achieves exact matches (0-frame error) for 86\% of user interactions. Furthermore, 89\% of interactions are measured within a 6-frame margin. 

Compared to response time, we see a decline in measuring the finish time. One cause of error is the presence of animations during or shortly after the user interaction. For example, some apps display animated content, such as dynamic advertisements or banners, after user interactions. These animations introduce abrupt visual changes, which our approach may mistakenly interpret as part of the interaction response, leading to a few extra frames being included in the measured finish time. However, since the updates of such animated content occur suddenly and only occasionally, they typically cause small errors (within a few frames).

\begin{table}[t]
    \centering
    \caption{Mean absolute differences between the measured and actual \textit{response time} of successfully identified interactions.}
    \label{tab:response_time_errors}
    
    \begin{minipage}{\columnwidth}
        \centering
        \textbf{(a) Frame-level differences}
        
        \begin{tabular}{l|c|c|c|c|c}
        \toprule
        MAE (\#frames) & =0 & $\le$1 & $\le$2 & $\le$3 & $>$3 \\
        \midrule
        1.2 & 92\% & 93\% & 94\% & 95\% & 5\% \\
        \bottomrule
        \end{tabular}
    \end{minipage}
    
    \vspace{1em}

    \begin{minipage}{\columnwidth}
        \centering
        \textbf{(b) Millisecond-level differences}

        \begin{tabular}{l|c|c|c|c|c}
        \toprule
        MAE (ms) & =0\,ms & $\le$17\,ms & $\le$33\,ms & $\le$50\,ms & $>$50\,ms \\
        \midrule
        33 & 92\% & 93\% & 93\% & 94\% & 6\% \\
        \bottomrule
        \end{tabular}
    \end{minipage}
\end{table}

\begin{table}[t]
    \centering
    \caption{Mean absolute differences between the measured and actual \textit{finish time} of successfully identified interactions.}
    \label{tab:finish_time_errors}
    
    \begin{minipage}{\linewidth}
        \centering
        \textbf{(a) Frame-level differences}

        \begin{tabular}{l|c|c|c|c|c}
        \toprule
        MAE (\#frames) & =0 & $\le$1 & $\le$3 & $\le$6 & $>$6 \\
        \midrule
        5.2 & 86\% & 87\% & 87\% & 89\% & 11\% \\
        \bottomrule
        \end{tabular}
    \end{minipage}
    
    \vspace{1em}

    \begin{minipage}{\linewidth}
        \centering
        \textbf{(b) Millisecond-level differences}

        \begin{tabular}{l|c|c|c|c|c}
        \toprule
        MAE (ms) & =0\,ms & $\le$17\,ms & $\le$50\,ms & $\le$100\,ms & $>$100\,ms \\
        \midrule
        198 & 86\% & 86\% & 87\% & 88\% & 12\% \\
        \bottomrule
        \end{tabular}
    \end{minipage}
\end{table}

We further observe that in 11\% of interactions, the finish time measured by \tool exceeds by six frames (i.e., 100 ms) or more. This is mainly due to errors in user interaction segmentation. When the next user interaction is missed in the segmentation, \tool mistakenly merges two interactions into one. As a result, it may select the finish frame of the second interaction as that of the first, leading to a large overestimation.
Nevertheless, such cases are relatively infrequent and do not significantly impact the overall performance of \tool. Its high accuracy across the majority of interactions demonstrates its effectiveness and practical usages in real-world settings.

\rqbox{\tool accurately measures response time for 92\% of user interactions and 95\% within three frames ($\approx$\!\,50 ms). For finish time, 86\% of interactions are measured correctly and 89\% fall within six frames ($\approx$\!\,100 ms).}

%% file: texFiles/rq3.tex
\begin{table*}[t]
\centering
\caption{Response time and finish time measurement accuracy across 32 app categories. App categories are sorted by the proportion of interactions with response time errors $\leq$3 frames (left) and finish time errors $\leq$6 frames (right).}
\label{tab:category_errors}
\setlength{\tabcolsep}{4pt}
\renewcommand{\arraystretch}{1.0}
\scalebox{0.95}{
\begin{tabular}{p{0.14\linewidth}cc|p{0.16\linewidth}cc||p{0.14\linewidth}cc|p{0.16\linewidth}cc}
\toprule
\multicolumn{6}{c||}{\textbf{Response Time Accuracy (Sorted by $\leq$3 frames)}} & \multicolumn{6}{c}{\textbf{Finish Time Accuracy (Sorted by $\leq$6 frames)}} \\
\textbf{App Category} & =0 & $\leq$3 & \textbf{App Category} & =0 & $\leq$3 &
\textbf{App Category} & =0 & $\leq$6 & \textbf{App Category} & =0 & $\leq$6 \\
\midrule
Books \& Reference       & 100.0 & 100.0 & Tools                   & 94.3 & 94.3 & Productivity             & 92.5 & 96.2 & Food \& Drink           & 84.6 & 88.5 \\
Education                & 100.0 & 100.0 & Social                  & 90.4 & 94.2 & Tools                   & 94.3 & 96.2 & Photography             & 84.5 & 88.1 \\
Travel \& Local          &  95.9 & 100.0 & Lifestyle               & 93.9 & 93.9 & Lifestyle               & 95.1 & 95.1 & News \& Magazines       & 79.3 & 87.9 \\
Shopping                 &  97.9 &  99.0 & Entertainment           & 92.4 & 93.7 & Education               & 92.6 & 94.7 & Travel \& Local         & 85.7 & 87.8 \\
Events                   &  95.1 &  98.8 & News \& Magazines       & 91.4 & 93.1 & Events                  & 91.5 & 93.9 & Comics                  & 84.8 & 87.3 \\
Productivity             &  97.5 &  98.8 & Photography             & 90.5 & 92.9 & Shopping                & 89.6 & 93.8 & Weather                 & 84.8 & 87.0 \\
Beauty                   &  98.2 &  98.2 & Art \& Design           & 87.3 & 92.1 & Finance                 & 93.5 & 93.5 & Auto \& Vehicles        & 86.9 & 86.9 \\
Communication            &  96.0 &  97.3 & Sports                  & 86.7 & 91.7 & Maps \& Navigation      & 93.4 & 93.4 & Personalization         & 82.2 & 86.7 \\
Libraries \& Demo        &  97.3 &  97.3 & Personalization         & 91.1 & 91.1 & Parenting               & 92.5 & 92.5 & Sports                  & 81.7 & 86.7 \\
Business                 &  95.7 &  96.8 & Medical                 & 81.5 & 90.2 & Communication           & 90.7 & 92.0 & Libraries \& Demo       & 79.1 & 82.7 \\
Music \& Audio           &  94.7 &  96.5 & House \& Home           & 86.7 & 90.0 & Books \& Reference      & 88.7 & 91.9 & Social                  & 82.7 & 82.7 \\
Finance                  &  95.3 &  96.3 & Comics                  & 88.6 & 89.9 & Entertainment           & 89.9 & 91.1 & Video Players \& Editors & 75.0 & 81.2 \\
Parenting                &  96.2 &  96.2 & Weather                 & 84.8 & 89.1 & Dating                  & 85.3 & 90.7 & Health \& Fitness       & 77.0 & 81.1 \\
Dating                   &  94.7 &  96.0 & Food \& Drink           & 79.5 & 88.5 & Medical                 & 85.9 & 90.2 & Beauty                  & 64.3 & 75.0 \\
Auto \& Vehicles         &  93.4 &  95.1 & Health \& Fitness       & 85.1 & 87.8 & Music \& Audio          & 84.2 & 89.5 & House \& Home           & 70.0 & 75.0 \\
Maps \& Navigation       &  90.8 &  94.7 & Video Players \& Editors & 87.5 & 87.5 & Business                & 88.2 & 89.2 & Art \& Design           & 69.8 & 74.6 \\
\midrule
\multicolumn{6}{c||}{\textbf{Avg Response Time Accuracy:} 92.2\% (=0-frame), 94.4\% ($\leq$3 frames)} & \multicolumn{6}{c}{\textbf{Avg Finish Time Accuracy:} 85.0\% (=0-frame), 88.2\% ($\leq$6 frames)} \\
\bottomrule
\end{tabular}
}
\end{table*}

\subsection*{RQ3: What is the Accuracy of GUI Responsiveness Measurement for Apps in Different Categories?}
Given the substantial variation in design and functional behavior across mobile apps, this RQ further assesses the measurement accuracy of \tool across 32 app categories in our dataset.
For each category, we examine the proportion of user interactions for which the response/finish time is measured with zero-frame error (matches the ground truth) and within a three-frame error. 

\phead{Results.} Table~\ref{tab:category_errors} presents the response time and finish time measurement accuracy across 32 app categories. Here, \textbf{accuracy} refers to the proportion of interactions whose measured time falls within a specified frame error bound (e.g., $\leq$3 frames for response time, $\leq$6 frames for finish time). 
Across all categories, \tool achieves 94.4\% accuracy within three frames for response time. For finish time, 85\% of interactions have zero-frame error, and 88.2\% fall within six frames. 
We sort the app categories based on the accuracy. We find that apps with more static content (e.g., those related to Books, Education, and Shopping) tend to have higher accuracy. In contrast, apps with more visual UI or animations (e.g., Video Players) tend to have a lower accuracy.

The difference in category rankings between response time and finish time is due to the different main factors that affect their measurement. For response time, lower-ranked app categories often contain reactive UI animations (e.g., a button dimming) that may be mistakenly identified as valid responses. In contrast, finish time is influenced by a wider range of factors. Since the finish frame comes after the response frame in the interaction, it is more likely to be affected by tap indicator misdetections between the actual response and finish frames.
In addition, animations or video playback that continue after the finish frame can obscure the actual end of the finish. Due to these additional challenges, finish time measurement is slightly less accurate than response time measurement. 

Overall, these results suggest that \tool performs reliably across a wide range of app categories, with high accuracy even in visually complex or animation-heavy apps. This highlights its robustness in handling diverse UI designs and interaction styles commonly found in mobile applications.

\rqbox{\tool achieves reliable accuracy across diverse app categories. Across 32 app categories, 94.4\% of interactions are measured within three frames for response time, and 88.2\% within six frames for finish time. Accuracy remains high even for visually complex apps, demonstrating \tool's broad applicability.} 

%% file: texFiles/rq4.tex
\subsection*{RQ4: How Useful is \tool for Performance Alerting?}
In practice, developers and testers are typically less concerned with the exact response or finish time of each interaction (e.g., 50 ms vs. 55 ms), but more focused on whether it feels slow or unresponsive from the user's perspective. 
To approximate this perception, thresholds like 100\,ms (response time) and 1,000\,ms (finish time) are widely adopted in both HCI research and industry~\cite{1968_AFIPS_Response_time_in_man_computer, 1994_Usability_Engineering, 2015_High_Performance_Android, 2022_TSE_Survey_of_Performance_Optimization}. We investigate whether \tool can act as a performance alerting system by identifying interactions that exceed these thresholds. This formulation aims to determine whether a user interaction is slow enough to warrant attention, as commonly required in industrial testing workflows regarding performance alerting and issue diagnosis.

\phead{Approach.} We evaluate \tool using a threshold-based classification method to determine whether it can accurately identify user interactions that violate standard responsiveness thresholds. We conduct an end-to-end evaluation, where \tool first detects all user interactions and measures the response/finish time. For each interaction, we compare the response or finish time measured by \tool with the corresponding ground-truth value, and consider the prediction correct if both fall on the same side of the threshold (i.e., either above or below). We evaluate classification accuracy using standard metrics (precision, recall, and F1-score) under thresholds commonly adopted in practice~\cite{1968_AFIPS_Response_time_in_man_computer, 1994_Usability_Engineering, 2015_High_Performance_Android, 2022_TSE_Survey_of_Performance_Optimization}: 100\,ms for response time and 1,000\,ms for finish time, where interactions above these values are likely to feel unresponsive and warrant further investigation.

\phead{Results.}
As shown in Table~\ref{tab:fixed_threshold_results}, 
\tool achieves a precision of 96.8\% and a recall of 96.3\% for the response time threshold (>100\,ms). For the finish time threshold (>1000\,ms), the precision and recall are 88.1\% and 90.6\%, respectively. 
We find that the results are not really impacted by the cumulative errors from the two-step process: 1) detecting the user interactions and 2) measuring response/finish time. 
In short, \tool achieves consistently high recall across both metrics, effectively capturing the majority of unresponsive interactions. This behavior aligns well with the needs of our industry partner, as we aim to identify as many potential issues as possible that require further investigation.

\rqbox{\tool enables reliable performance alerting, achieving over 96\% precision and recall for detecting slow response times, and maintains strong recall (90.6\%) with reasonable precision (88.1\%) for finish time issues. Its high recall helps flag most unresponsive interactions, supporting early detection and industrial prioritization.}

\begin{table}[t]
\centering
\caption{Detection results of \tool for unresponsive interactions under fixed thresholds. }
\label{tab:fixed_threshold_results}
\setlength{\tabcolsep}{3pt}
\begin{tabular}{l|c|c|c|c|c}
\toprule
\textbf{Metric} & \shortstack{\textbf{Threshold} \\ \textbf{(ms)}} & \shortstack{\textbf{Interaction} \\ \textbf{Count}} & \shortstack{\textbf{Precision} \\ \textbf{(\%)}} & \shortstack{\textbf{Recall} \\ \textbf{(\%)}} & \shortstack{\textbf{F1-score} \\ \textbf{(\%)}} \\
\midrule
Response Time & $>100$  & 1299 & 96.8 & 96.3 & 96.5 \\
Finish Time   & $>1000$ & 508 & 88.1 & 90.6 & 89.3 \\
\bottomrule
\end{tabular}
\end{table}

%% file: texFiles/rq5.tex
\subsection*{RQ5: What is the Efficiency and Recording Overhead of \tool?}
\label{sec:rq5}

We evaluate two practical aspects of \tool: (1) the efficiency of processing recorded video, since it needs to analyze thousands of videos daily, and (2) the performance overhead introduced by screen recording, as it may affect the measurement results. 

\phead{Approach.} 1) Video Processing Efficiency. \tool leverages computer vision to analyze mobile-recorded video and measure responsiveness metrics. We measured the runtime performance on a workstation equipped with an NVIDIA RTX 4090 GPU, a 16-core AMD CPU, and 64 GB of RAM. We report the average processing time per user interaction, which includes the time for segmenting videos into user interactions and measuring responsiveness metrics. Since the length of each interaction varies, we also report the minimum and maximum processing times to highlight the range. 

2) Recording Overhead. Recording the screen of mobile apps may impact app responsiveness. To evaluate this, we conducted a controlled experiment using a Google Pixel 7 smartphone running Android 14, testing two configurations. For \textbf{No Recording (Baseline)}, we use an external camera to record the mobile screen with tap indicator disabled. We infer the user operations by manually observing the moment when the user's finger physically touches the screen. This setup avoids any software-based recording overhead while still allowing us to annotate response time and finish time based on visible UI changes. When using \textbf{scrcpy} (recording framework used in \tool), we enable screen recording with tap indicator turned on. 

Since the goal of \tool is to measure the human-perceived GUI responsiveness, we conduct the performance measurement manually. We randomly selected three mobile apps from the top apps. For each app, we identified 10 distinct user interactions and repeated each interaction 10 times using Pixel 7. For each interaction in the recorded video, we manually annotated the response and finish times across the 10 repetitions, and computed their average to obtain a representative value. We repeat the process for the baseline and for using \texttt{scrcpy}. To enable fair comparison, we compared the same interaction across different recording settings. We computed the \textbf{$\Delta$ Response Time (ms)} and \textbf{$\Delta$ Finish Time (ms)}, i.e., the difference between each recorded configuration and the baseline, for each corresponding interaction. 

\phead{Results.} \tool processes each frame in approximately 30\,ms on average, and its total processing time scales linearly with the number of frames. For instance, a 5-second user interaction recorded at 60\,fps (300 frames) requires about 9.0\,s to process. As also observed with our industry partner, \tool can process thousands of recordings within an hour. 
In terms of recording overhead, using \texttt{scrcpy} introduces only a small average error of $\Delta$16\,ms in response time and $\Delta$51\,ms in finish time. These results suggest that 
\texttt{scrcpy}-based recording is efficient and adds minimal performance overhead, which is mostly unnoticeable by humans~\cite{1968_AFIPS_Response_time_in_man_computer, 1994_Usability_Engineering, 2015_High_Performance_Android, 2022_TSE_Survey_of_Performance_Optimization}, making it suitable for measuring GUI responsiveness.

\rqbox{\tool processes each frame in $\sim$30\,ms (about 9\,s for a 5-second interaction at 60\,fps), and introduces minimal recording overhead ($\Delta$16\,ms response, $\Delta$51\,ms finish), making it efficient and practical for analyzing GUI responsiveness.} 

%% file: texFiles/discussion.tex
\section{Production Deployment and Feedback from Our Industry Partner}
\label{sec:discussion}

In this section, we discuss the feedback that we received from our industry partner regarding the use of \tool in production testing environments. 

\subsection{Integration into Industrial Testing Pipeline}
\tool is deployed in an industrial setting and integrated into their existing automated mobile testing pipeline. The pipeline executes thousands of automated GUI tests across hundreds of mobile apps daily. Every test is recorded in high resolution with visible tap indicators turned on. \tool processes these screencasts in parallel, automatically identifying user interactions, measuring response and finish times, and flagging interactions that exceed user-perceived latency thresholds. Since \tool required no changes to the test scripts or application binaries, the tool integration was seamless. 
This ease of adoption was particularly appreciated, as modifying app code or injecting instrumentation is often infeasible in large-scale testing.

\subsection{Values in Uncovering User-Perceived Issues} 
A major benefit reported by the development team was \tool's ability to detect GUI responsiveness issues that other tools failed to identify. Conventional approaches, such as performance profilers, execution traces, and log-based analysis, rely on system-level metrics that often do not correlate with what users perceive and experience. As a result, subtle but perceptible delays were often overlooked. 
The feedback we received was that \tool, which analyzes screencasts to flag interactions that appear visually delayed, provides them a better understanding of the user-perceived performance issues. In several instances, the tool uncovered problems that had been previously flagged by users in negative app reviews, but which had not been reproducible or diagnosable using internal tools. 
These findings highlighted the gap between traditional performance monitoring and user-perceived responsiveness, and how approaches like \tool can fill this gap.

\subsection{Actionable Feedback and Debugging Support} 
\tool also helped improve the developer experience during performance debugging. The tool outputs precise timestamps and specific start/end frames for each interaction. It also provides annotated video segments that exceed responsiveness thresholds. Developers found this output intuitive and actionable, where they can visually inspect a small video segment and immediately understand when and where the delay occurred. 
For instance, one developer mentioned that ``Before \tool, we had to guess whether a delay reported by QA was real or perceptible. Now, we just click the video segment and see the problem unfold frame by frame.'' 
In summary, the positive feedback highlights the benefits of \tool in debugging support and how it provides a different perspective of performance debugging.

%% file: texFiles/threats.tex
\section{Threats to Validity}
\label{sec:threats}
\phead{External Validity.}
One threat to external validity is the generalizability of our evaluation dataset. While it includes 64 apps from 32 diverse categories, the dataset may not fully represent the wide range of mobile app designs and usage scenarios in practice. 
Future studies may evaluate \tool on a broader range of apps, including a wider variety of interaction styles.

\phead{Construct Validity.}
A major threat to construct validity is human error in the manual video annotation process in establishing the ground truth. To mitigate the risk, three experienced authors independently labeled the start, response, and finish frames for over 2,400 user interactions. Disagreements were resolved through discussion and consensus. 

\phead{Internal Validity.}
Screencast recording and enabling the tap indicator may introduce performance overhead that could influence the internal validity of the results. However, our findings in RQ5 show that the overhead is minimal and mostly imperceptible to users. 

%% file: texFiles/conclusion.tex
\section{Conclusion}\label{sec:conclusion} 
In this paper, we presented \tool, a practical framework for measuring GUI responsiveness in mobile applications using recorded screencasts. 

Our evaluation shows that \tool achieves high precision and recall in detecting interactions and provides accurate responsiveness measurements across diverse app categories. When used as a performance alerting system, it effectively flags interactions that exceed industry-relevant thresholds. We further deployed \tool in an industrial setting, where it received positive feedback for uncovering user-perceived issues that existing tools failed to detect.

Overall, \tool offers a practical and effective solution for detecting GUI unresponsiveness in mobile apps by integrating computer vision techniques into performance analysis.